\documentclass[pra,twocolumn,preprintnumbers,superscriptaddress]{revtex4-1} 

\usepackage{latexsym,epsfig,amssymb,amsfonts,amsmath,graphicx,bbm,bm}

\usepackage{color}
\definecolor{dred}{rgb}{.8,0.2,.2}
\definecolor{dyellow}{rgb}{.7,.7,.0}
\definecolor{ddred}{rgb}{.4,.0,.0}
\definecolor{dblue}{rgb}{.2,.2,.8}
 
\newcommand*{\rom}[1]{\uppercase\expandafter{\romannumeral #1\relax}}

\newcommand{\dd}{\; \mathrm{d}}

\newcommand{\pd}[2]{\frac{\partial #1}{\partial #2}}

\newcommand{\eqr}[1]{Eq.\ (\ref{#1})}
\newcommand{\fir}[1]{Fig.\ \ref{#1}}

\newcommand{\ee}{\textrm{e}}

\newcommand{\iii}{\mathbf{i}}

\newcommand{\zz}{\mathbf{z}}

\newcommand{\OO}{\mathcal{O}}

\newcommand{\EE}{\mathcal{E}}

\begin{document}

\title{Capturing exponential variance using polynomial resources: applying tensor networks to non-equilibrium stochastic processes}

\author{T. H. Johnson}
\email{tomihjohnson@gmail.com} 
\affiliation{Centre for Quantum Technologies, National University of Singapore, 3 Science Drive 2, 117543, Singapore}
\affiliation{Clarendon Laboratory, University of Oxford, Parks Road, Oxford OX1 3PU, United Kingdom}
\affiliation{Keble College, University of Oxford, Parks Road, Oxford OX1 3PG, United Kingdom}
\affiliation{Institute for Scientific Interchange, Via Alassio 11/c, 10126 Torino, Italy}
\author{T. J. Elliott}
\affiliation{Clarendon Laboratory, University of Oxford, Parks Road, Oxford OX1 3PU, United Kingdom}
\author{S. R. Clark}
\affiliation{Clarendon Laboratory, University of Oxford, Parks Road, Oxford OX1 3PU, United Kingdom}
\affiliation{Centre for Quantum Technologies, National University of Singapore, 3 Science Drive 2, 117543, Singapore}
\affiliation{Keble College, University of Oxford, Parks Road, Oxford OX1 3PG, United Kingdom}
\author{D. Jaksch}
\affiliation{Clarendon Laboratory, University of Oxford, Parks Road, Oxford OX1 3PU, United Kingdom}
\affiliation{Centre for Quantum Technologies, National University of Singapore, 3 Science Drive 2, 117543, Singapore}
\affiliation{Keble College, University of Oxford, Parks Road, Oxford OX1 3PG, United Kingdom}

\date{\today}

\begin{abstract}
Estimating the expected value of an observable appearing in a non-equilibrium stochastic process usually involves sampling. If the observable's variance is high, many samples are required. In contrast, we show that performing the same task without sampling, using tensor network compression, efficiently captures high variances in systems of various geometries and dimensions. We provide examples for which matching the accuracy of our efficient method would require a sample size scaling exponentially with system size. In particular, the high-variance observable $\ee^{-\beta W}$, motivated by Jarzynski's equality, with $W$ the work done quenching from equilibrium at inverse temperature $\beta$, is exactly and efficiently captured by tensor networks.
\end{abstract}


\maketitle 
{\em Introduction.---}Dynamical stochastic processes are used throughout the natural and social sciences when inaccessible degrees of freedom are well-represented by random variables~\cite{VanKampen2007,Rolski1999}. To calculate expected observable values, numerical methods are usually required. Out of equilibrium, the typical method is dynamical Monte Carlo~\cite{Cox1965,Young1966,Bortz1975,Gillespie1976,Jensen2003,Chatterjee2007}. Essentially, averaging over randomly sampled paths provides an unbiased estimate for the expected value of an observable. To obtain a fixed expected fractional error, the number of paths sampled must scale linearly with the variance divided by the square of the expected value. For a multitude of important observables, such as those appearing in the estimation of free energies~\cite{Potamianos1997} and likelihoods of rare events~\cite{Dellago1998,Mazonka1998,Grassberger2002,Lee2005,Dean2009,Kundu2011}, this ratio is large: such observables are said to have high variance and sampling methods struggle when applied to them.

Here we present an approach that is very different to sampling. We simultaneously follow all paths, which is made efficient by using controlled data compression, usually approximate but exact in special cases, based on tensor networks.
While tensor networks have previously been used in conjunction with stochastic processes~\cite{Hieida1998,Carlon1999,Carlon2001,Temme2010,Johnson2010}, the question of how their performance relates to variance has remained unanswered. Understanding this is crucial if we are to know whether or not tensor networks, which have had a revolutionary effect in simulating quantum systems~\cite{Verstraete2009,Cirac2009,Schollwock2011,AlAssam2011,Denny2011,Evenbly2013,Orus2013}  and have been used to great effect in solving partial differential equations~\cite{Iblisdir2007,Koch2007,Holtz2012,Lubich2013}, provide a useful and perhaps essential complementary technique to sampling in stochastic processes. 

In this Letter we address this question and our answer is very clear: high variance does not limit the accuracy of tensor network compression, and tensor networks can be applied efficiently to tackle problems, even out of equilibrium, for which sampling-based methods struggle.  
This opens the door for the use of tensor network methods on a wide-variety of non-equilibrium stochastic systems for which capturing high variance is essential. In particular, we show that a distribution of weighted expectation values of high-variance observable $\ee^{-\beta W}$, with $W$ the work done quenching from equilibrium at inverse temperature $\beta$, is represented exactly by a highly compressed tensor network.    


\begin{figure}[t]
\includegraphics[width=8.6cm]{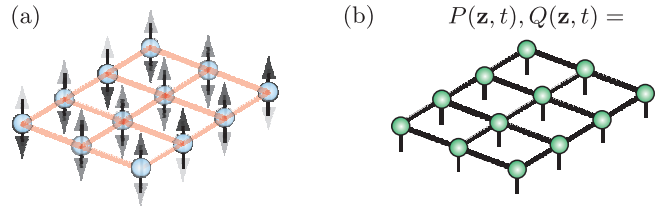}
\caption{\label{fig:tensornetwork} (color online) {\em Tensor network compression}. (a) An Ising system whose degrees of freedom (blue circles with arrows) interact, in this case, with a two-dimensional lattice geometry (red lines). (b) The probability distributions $P_E(\zz)$ and $P(\zz,t)$, and $Q(\zz,t)$ (see main text) at any time $t$ are compressed by representing them (approximately or, in special cases, exactly) by a contraction of tensors (green circles) with the same geometry as the interactions. Each black leg corresponds to an index of a tensor, and the joining of two legs represents the contraction of the two corresponding indices. 
}
\end{figure}

We focus on an Ising system, an example of which is shown in \fir{fig:tensornetwork}(a). It comprises $N$ nodes, labeled by $\ell$, the configuration $z_\ell$ of each taking one of $d=2$ discrete values $z_\ell = \{-1,1\}$ . The configuration of all $N$ nodes is given by the $N$-tuple $\zz = ( z_1, \dots , z_N)$ and the probability of being in configuration $\zz$ is $P(\zz)$. Tensor networks  best suit systems for which crucial quantities, like energy, are $n$-bodied, with $n$ small. We consider the simplest non-trivial case of an energy comprising single and two-body terms
\begin{equation}
\label{eq:Ising}
E (\zz) = - J  \sum_{( \ell, \ell' ) \in \EE} z_\ell z_{\ell'} - \lambda \sum_\ell  z_\ell \, ,
\end{equation}
where $\EE$ are $N_\EE$ edges connecting interacting nodes.

{\em Equilibrium.---}The relationship between compressibility and variance out of equilibrium builds on that in equilibrium. 
The equilibrium Gibbs distribution at inverse temperature $\beta$ is $P_E  (\zz) = \ee^{-\beta E (\zz)}$, normalized to the partition function $Z_E = \sum_\zz \ee^{-\beta E (\zz)}$. 
It is always possible to represent $P_E  (\zz)$ (or any distribution) by a tensor network of the form shown in \fir{fig:tensornetwork}(b)
\begin{equation}
\label{eq:tn}
P_E  (\zz) = \sum_{\iii} \prod_{\ell} A^{[\ell] z_\ell}_{\iii_{\ell}},
\end{equation}
that shares the same geometry as the interactions.
Here $A^{[\ell]}$ is a tensor associated with node $\ell$. It has a physical index $z_\ell$ and $k_\ell$ auxiliary indices $\iii_{\ell}$, one for each edge connected to $\ell$, and each taking one of $\chi$ values, which may in principle be large. The sum is over the values taken by all auxiliary indices,
which is just a sum over an $N_\EE$-tuple $\iii$ of indices. The Gibbs distribution is important because the tensor network representation is exact for $\chi =  d$ (see Supplemental Material \cite{SM}). This implies that $\sum_\ell d^{k_\ell+1}$ numbers may be used to represent $d^N$ others, providing a significant yet exact compression if the degrees $k_\ell$ are limited, as in lattice systems with local interactions.

As well as compression, tensor networks offer a means of calculating the partition function $Z_E$, since $Z_E = \sum_\zz P_E  (\zz) = \sum_{\iii} \prod_{\ell} T^{[\ell] }_{\iii_{\ell}}$ with transfer tensors $T^{[\ell] } = \sum_{z_\ell} A^{[\ell] z_\ell}$~\cite{Nishino1995,Nishino1996,Levin2007,Orus2008}. For a one-dimensional (1D) chain
  the partition function $Z_E$ relates to a product of transfer matrices and requires $\OO(N d^2)$  or $\OO(N d^3)$ resources to compute for open or periodic boundaries, respectively. In higher dimensions (if the tensor network has a large treewidth~\cite{Shi2006,Markov2008}) the tensor contractions cannot in general be performed both exactly and efficiently, but efficient strategies exist to perform them approximately. Levin and Nave~\cite{Levin2007} demonstrated that this can be done accurately for two-dimensional (2D) non-critical lattice systems using tensor renormalization group~\cite{Xie2009,Zhao2010}.

Contrastingly, estimating the partition function $Z_E$ directly by evaluating the sum $Z_E = \sum_\zz P_E  (\zz)$ through random sampling is made difficult by the fact that, in general, the variance of observables requiring estimation grows quickly with system size~\cite{Potamianos1997}. The tensor network representations of $P_E (\zz)$ and $Z_E$ show that high variance does not imply difficulty in equilibrium, away from criticality. 

{\em Non-equilibrium.---}Out of equilibrium, the dynamics of a Markovian system~\cite{Norris1998} depends only on its current configuration, and the evolution of the distribution $P(\zz,t)$ is described by a master equation of the form
\begin{equation}
\label{eq:Markovian}
\pd{P (\zz, t)}{t} = \sum_{\zz'} H (\zz,\zz',t) P (\zz', t) .
\end{equation}
Each non-negative off-diagonal element $H (\zz,\zz',t)$ for $\zz \neq \zz'$ is the Poisson rate of a transition from $\zz'$ to $\zz$ at time $t$, and together these fix the non-positive diagonals $H (\zz,\zz,t) = - \sum_{\zz' \neq \zz} H (\zz',\zz,t)$ such that the normalization of $P(\zz,t)$ is conserved. $H$ is commonly referred to as the Hamiltonian.

To simulate such dynamics using non-equilibrium tensor network methods, we represent $P (\zz, t)$ at any time by a tensor network, as in \eqr{eq:tn}, with a small $\chi$. Doing so assumes that this representation, while not necessarily exact, is accurate. There is no guarantee of this accurate compressibility on all occasions, but it is expected in many situations. For example, consider a quench from one Hamiltonian $H (\zz,\zz',0) = H_0 (\zz,\zz')$ to another $H (\zz,\zz',\tau) = H_1 (\zz,\zz') \neq H_0 (\zz,\zz')$, where the system begins in the stationary state $P_0 (\zz)$ satisfying $\sum_{\zz'} H_0 (\zz,\zz') P_0 (\zz') =0$. For much later times $t \gg \tau$ (on the timescale $h^{-1}$, where $h$ is some Hamiltonian-specific convergence rate), the system will converge to another stationary state $P_1 (\zz)$ satisfying $\sum_{\zz'} H_1 (\zz,\zz') P_1 (\zz') =0$. Numerous examples have revealed that stationary states of local stochastic processes are accurately compressible via tensor network representations~\cite{Hieida1998,Carlon1999,Carlon2001,Temme2010,Johnson2010}. Thus in such quenches both initial and long-time distributions $P_{0} (\zz)$ and $P_{1} (\zz)$ are accurately compressible. Unlike for quantum systems~\cite{Calabrese2005}, compression errors are limited even when a system is driven away from equilibrium. The example on which we focus here is that of a thermalizing (equilibrating) Hamiltonian $H (\zz,\zz',t) = H_E (\zz,\zz')$ for which the Gibbs distribution is stationary,
$\sum_{\zz'} H_E (\zz,\zz') P_E (\zz') =0$,
and it is the energy $E(t)$ that is quenched by varying the bias $\lambda (t)$ appearing in \eqr{eq:Ising}. 

The probability distribution $P(\zz,t)$ over configurations contains only partial information about the full probability distribution over the possible paths through configuration space taken by the stochastic process. As such, the expected values of only certain observables may be calculated from $P(\zz,t)$. These include observables whose values $O(\zz)$ depend on the configuration $\zz$ of the system at a single time $t$, thus having expected value $\langle O(t) \rangle = \sum_\zz O(\zz) P(\zz,t)$. We call such observables configuration-dependent and use the example of the magnetization $M (\zz) = \sum_\ell z_\ell $. The values of some other observables depend on the full path taken by the system and their expected values cannot be calculated from $P(\zz,t)$. We call such observables path-dependent, and use the example of the work done $W(t) = -\int_0^t \dd s M(\zz(s)) \dot{\lambda}(s)$ by varying $\lambda(t)$ between times $0$ and $t$. 

Although not previously considered in the literature, the expected values of some path-dependent observables can indeed be calculated using tensor networks and, as we will show, provide us with a stark example of exact compressibility in the face of high variance out of equilibrium. The idea is to represent the relevant path-dependent information locally in time, not with $P(\zz ,t)$, but through the distribution of weighted conditional expected values $Q (\zz , t) = P(\zz, t) \langle O (\zz,t) \rangle$. Here $\langle O (\zz,t) \rangle$ is the expected value an observable has accumulated by time $t$ conditioned upon the system arriving at configuration $\zz$ at that time. The expected value of interest $\left \langle O(t) \right \rangle = \sum_\zz Q (\zz , t)$ is then obtained from $Q (\zz , t)$.
It follows from \eqr{eq:Markovian} that the distribution $Q(\zz, t)$ evolves as
\begin{align}
\label{eq:evolution}
\pd{Q (\zz, t)}{t} &= \sum_{\zz'} H'  (\zz,\zz',t) Q (\zz', t) ,
\end{align}
with $H'  (\zz,\zz',t) = H  (\zz,\zz',t) + \dot{o} ( \zz , t) \delta (\zz, \zz' )$, where $\dot{o} (\zz , t)$ is the rate of increase of the natural logarithm of the observable at configuration $\zz$ and time $t$. 

It is desirable to predict, as we have for $P (\zz , t)$, the accuracy of compressing $Q (\zz , t)$ at any time during its evolution using tensor networks. 
Consider the quench in $\lambda (t)$ between times $0$ and $\tau$, starting from equilibrium. 
The distributions are initially equal $Q (\zz , 0) = P (\zz , 0)$, thus the accurate compressibility of the latter implies the same of the former. Additionally, $\dot{o} (\zz , t)$ is only non-zero for times $t<\tau$ and the stochastic evolution is ergodic. Thus after a sufficiently long time $t\gg \tau$ (relative again to convergence timescale $h^{-1}$) the configurations will have mixed such that $\langle O (\zz,t) \rangle = \langle O (t) \rangle$ is independent of $\zz$ and thus once again $Q (\zz , t) = \langle O (t) \rangle P (\zz , t)$ is as accurately compressible as $P (\zz , t)$. 


\begin{figure}[tbp]
\includegraphics[width=8.6cm]{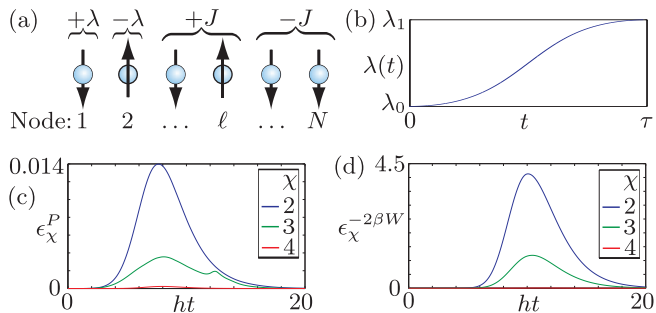}
\caption{\label{fig:introfig} (color online) {\em Accurate compression out of equilibrium}. (a) An $N$-node open Ising chain with exchange $J$ and bias $\lambda$. (b) The bias $\lambda (t)$ is  varied, driving the system away from equilibrium. (c) The fractional error $\epsilon_1$ between calculating $\langle \ee^{M(t)} \rangle$ using probability distribution $P(\zz,t)$ and its compressed tensor network approximation of dimension $\chi$ (see Supplemental Material~\cite{SM}). A dashed line marks the end of the quench at time $t = \tau$. (d) Similarly, the fractional error $\epsilon_2$ between calculating $\langle \ee^{-2\beta W(t)} \rangle$ by summing $Q(\zz,t)$ and its compressed tensor network approximation. The parameters used are $\beta \lambda_0 = 0$, $\beta \lambda_1 = 1$, $h\tau = 10$, $\beta J = 1$ and $N=8$.
}
\end{figure}

{\em Numerical examples.---}We demonstrate these behaviors for a system undergoing thermalizing Glauber dynamics~\cite{Glauber1963} via local transitions,
$$H_E  (\zz,\zz')  = h(\zz,\zz') \left [1 + \ee^{-\beta \left( E (\zz') - E (\zz)  \right) } \right]^{-1}, $$ 
for $\zz \neq \zz'$. Here $h(\zz,\zz') = h(\zz',\zz)$ are symmetric rates equaling a non-zero rate $h$ only where $\zz$ and $\zz'$ differ by the configuration of a single node. The energy is quenched via the parameter $\lambda (t)$ varying from $\lambda_0$ to $\lambda_1$ over time $0 \leq t \leq \tau$ according to a smoothed tanh ramp (see Supplemental Material~\cite{SM}), as drawn in \fir{fig:introfig}(b). We focus on configuration-dependent observable $\ee^{M(t)}$ and path-dependent observables $\ee^{-2\beta W(t)}$ and $\ee^{-\beta W(t)}$. All have variance over mean squared growing exponentially with system size $N$. Initially, we consider the Ising nodes to be in an open 1D chain, illustrated in \fir{fig:introfig}(a).

To assess the accuracy of compression, we exactly calculate distributions $P(\zz,t)$ and $Q(\zz,t)$ for small $N=8$ at time $t$, then calculate the error in the expected value of observables induced by compressing the distributions as a tensor network. The errors shown in Figs.~\ref{fig:introfig}(c) and (d) for $\ee^{M(t)}$ and $\ee^{-2\beta W(t)}$, respectively, show that, despite large variances, expected values are relatively unaffected by tensor network compression. The distributions are exactly compressible at $t=0$ and thus no error occurs, as expected. The errors due compression initially increase as $\lambda (t)$ varies, then decrease exponentially to small values again on a timescale $\sim h^{-1}$. Interestingly, errors begin to decrease even at times $t<\tau$ prior to the end of the quench. For $\chi \gtrsim 4$ the compression is near-exact at all times. We arrive at similar conclusions for other types of variation tried e.g.\ linear, variations in $J$, and other observables.

A striking example is found in the path-dependent observable $\ee^{-\beta W(t)}$. The observable has received particular attention due to its featuring in several non-equilibrium identities in statistical physics, such as that by Jarzynski~\cite{Jarzynski1997}. Crucially for our discussion, the {\em non-equilibrium} distribution $Q (\zz , t) = P(\zz, t) \langle O (\zz,t) \rangle$ for this special case has an {\em equilibrium} structure
$Q (\zz , t) = P_{t} (\zz) Z_t(\beta)/ Z_0(\beta)$~\cite{Jarzynski1997},
where we have used shorthands of the form $ P_{t} (\zz)$ for the Gibbs distribution corresponding to $\lambda(t)$ and $Z_t(\beta)$ for the corresponding partition function (where from now on we normalize the Gibbs distribution to $1$). It immediately follows from our discussion of systems in equilibrium that $Q (\zz , t)$, despite containing information about non-equilibrium high-variance observables, has an exact highly-compressed $\chi = d$ tensor network representation at all times $t$. This can be used to efficiently and accurately calculate not only $\langle O(t) \rangle$ but a range of properties of the work distribution during such dynamics. Note that this exact behavior is particular to $\ee^{-\beta W(t)}$ and doesn't even extend to its square $\ee^{-2\beta W(t)}$.

We have so far demonstrated accurate single-time compressibility. We next examine how this extends to a dynamical tensor network simulation, where compression of $P (\zz , t)$ or $Q (\zz , t)$ occurs not only at a single time but at all times during their evolution.
While one might expect the compression errors at single times to accumulate, we find this is mitigated by the ergodicity of the evolution (unlike in quantum systems). For example, errors in $P(\zz,t)$ will not change the distribution to which it converges, and thus the significance of transient errors diminish, rather than accumulate, in time. In what follows, the specific algorithm we use to perform the evolutions of Eqs.\ (\ref{eq:Markovian}) and (\ref{eq:evolution}) is time-evolution block decimation (TEBD)~\cite{Vidal2004,Johnson2010,TNTLibrary}. The TEBD algorithm uses a timestep $\delta t$ resulting in an error, beyond that due to compression, of $\OO(N \delta t^2)$ and requires time $\OO (N \chi^{3} \delta t^{-1})$~(see the Supplemental Material~\cite{SM}).

\begin{figure}[tbp]
\includegraphics[width=8.6cm]{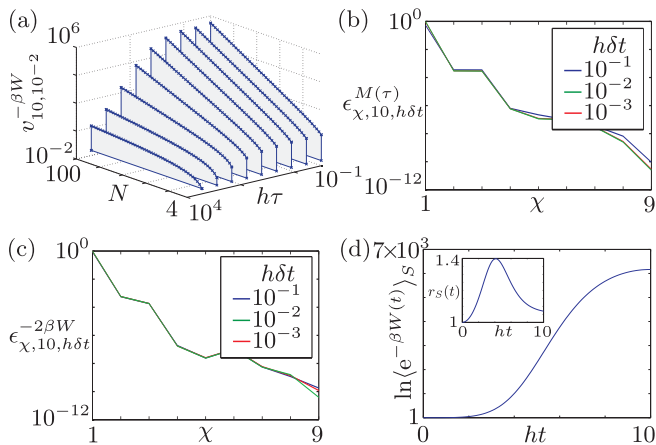}
\caption{\label{fig:Jarzynskiverification} (color online) {\em Compression and variance}. 
(a) The fractional error $\epsilon_1$ of the calculated expected value $ \langle \ee^{M} \rangle$ relative to the $\chi = 10$ result. (b) The analogous error $\epsilon_2$ for observable $\ee^{- 2 \beta W}$. (c) The variance over mean squared $v_3$ of $\ee^{-\beta W}$ as a function of $h \tau$ and $N$, obtained by calculating $\langle \ee^{-2\beta W} \rangle$ using $\chi = 10$ and $\delta t = 10^{-2}$, and using an exact result for $\langle \ee^{-\beta W} \rangle$. (d) For a periodic $N = 64 \times 64$ system, the expectation value $\langle \exp^{-\beta W(t)} \rangle_S$, conditional on one spin taking value $z_1 = 1$ at $t$ (full line). Also, the ratio $r_S (t)$ of this value relative to that for a reversible quench (dashed line). Unless stated otherwise the parameters used are $\beta \lambda_0 = 0$, $\beta \lambda_1 = 1$, $h\tau = 10$, $\beta J = 1$, $N=200$ and $h\delta t = 10^{-3}$.
}
\end{figure}

We first calculated $\langle \ee^{M} \rangle$ and $\langle \ee^{- 2\beta W} \rangle$, where $M = M(\tau)$ and $W = M(\tau)$ are values at the end of the quench $t = \tau$.
There are no exact values available to compare against for large systems, but the compressibility expected from Figs.~\ref{fig:introfig}(c) and (d) is confirmed by our TEBD results in Figs.~\ref{fig:Jarzynskiverification}(a) and (b), respectively. The calculated expected values converge approximately exponentially with increasing $\chi$, reaching acceptably converged values by $\chi \lesssim 5$, despite the variance over mean squared being very large for the $N=200$ considered.

We next calculated $ \langle \ee^{-\beta W} \rangle$. Since compression is exact for this observable if $\chi \geq 2$, the error $\epsilon_3$ is purely due to the finite timestep $\delta t$ and, as stated above, scales as $\epsilon_3 \propto N$. Meanwhile the estimated variance over mean squared $v_{3}$ scales with $\ee^{\OO (N)}$, as shown in \fir{fig:Jarzynskiverification}(c). It is then particularly clear with this example that, to achieve the same fractional accuracy (namely, $\epsilon_3 \propto N$) as we efficiently obtain, a na\"{\i}ve sampling method would require a sample size and thus time $\ee^{\OO(N)}$ in contrast to the TEBD algorithm that requires time $\OO(N)$. Explicitly, our $N=200$, $\chi = 2$, $h \delta t = 10^{-4}$, $h \tau =10$ calculation takes less than an hour and achieves an error $\epsilon_3 \approx 10^{-6}$. We do not compare this against the cpu time of any one sampling algorithm, as this choice is likely to be unrepresentative. Instead, we note that, since $v_3 \approx 10^{11}$, our accuracy would require $\approx10^{17}$ samples to reproduce, and so matching our cpu time would require each path to be sampled in $\approx10^{-13}\mathrm{s}$.

As our final example, we demonstrate both the diversity of information stored in the distribution $Q (\zz , t)$ and the application of our method to 2D lattice geometries, specifically a $N = 64 \times 64$ square periodic lattice. We consider $\langle O(t) \rangle_S = \sum_{\zz \in S} P (\zz , t) \langle O(\zz, t) \rangle / \sum_{\zz \in S} P(\zz,t)  $, the expected value of $O(t) = \ee^{-\beta W(t)}$ given the system's configuration $\zz$ is in subset $S$ at time $t$, where $S$ is the set configurations in which one node has value $z_1 = 1$. The observable has a very large variance. Its conditional expected value can be rewritten $\langle O(t) \rangle_S = \sum_{\zz \in S} Q (\zz , t) / \sum_{\zz \in S} P(\zz,t)$. The numerator $\sum_{\zz \in S} Q (\zz , t)$ corresponds to a high-variance observable but can nevertheless be efficiently evaluated using equilibrium techniques for Gibbs distributions, due to the equilibrium structure of the non-equilibrium distribution $Q (\zz , t)$. In this case we use the tensor renormalization group method~\cite{Levin2007} with intermediary dimension $\chi = 3$ (see Supplemental Material~\cite{SM}), which suffices as the model parameters are far from criticality. The denominator $\sum_{\zz \in S} P(\zz,t)$ is simply the low-variance expected value of an observable taking value unity when $\zz(t) \in S$, otherwise zero. This can be accurately calculated using our tensor network methods or the more common dynamical Monte Carlo methods~\cite{Bortz1975,Gillespie1976,Jensen2003}. We used the latter, with sample size $\sim 1.5 \times 10^4$ (see Supplemental Material~\cite{SM}). The result is plotted in Fig.~\ref{fig:Jarzynskiverification}(d). Also plotted is $r_S (t)$, the ratio of the expected value $\langle O(t) \rangle_S$ to its value were the system in equilibrium at all times. This emphasizes that, despite exploiting an equilibrium structure, the dynamics being simulated is truly irreversible.

{\em Discussion.---}We have shown that tensor networks provide a way to overcome the challenges faced by sampling methods when estimating expected values of high-variance observables out of equilibrium, even finding an exactly compressible non-equilibrium example. While advanced techniques for variance reduction exist, such as sequential importance sampling~\cite{Dellago1998,Mazonka1998,Lee2005,Kundu2011} and branching methods~\cite{Grassberger2002,Dean2009}, using these techniques usually requires judicious choices specific to the models to be simulated based on prior intuition about the process. No such intuition or choices are needed for a tensor network calculation. However, whether tensor networks remain accurate and efficient for geometries beyond 1D and 2D lattices, and other models e.g.\ describing frustration or disorder, must still be established.

Finally let us comment on how our findings relate to the wider use of tensor networks. 
While outstanding efficiency is possible using dynamical tensor network algorithms for finite 1D pure {\em quantum} systems, there is an ongoing effort from the community to reach larger dimensions $\chi$ and sizes $N$ in 2D. Reference \cite{Lubasch2014} gives a state-of-the-art demonstration in which the ground state of an $N=21 \times 21$ system with $d=2$ is calculated using $\chi = 8$. Meanwhile we have seen here that often very small dimensions $\chi$ are required to simulate {\em classical} systems, even during real time dynamics. 
Further, since only one copy of $P(\zz)$ is needed in classical algorithms, compared to two copies of the wavefunction in quantum algorithms, those that take time e.g. $\OO( \chi^6)$ for quantum systems will instead take time e.g. $\OO(\chi^3)$ for classical systems. It may therefore be the case that classical stochastic systems are currently in an even better position than quantum systems to benefit from current high-dimensional tensor network algorithms.

THJ, SRC, and DJ thank the National Research Foundation and the Ministry of Education of Singapore for support. TJE thanks St Anne's College of the University of Oxford for financial support. This research received funding from the ERC under the European Unionʼs Seventh Framework Programme (FP7/2007-2013)/ERC Grant Agreement no. 319286 Q-MAC and from the EPSRC through projects EP/K038311/1 and EP/J010529/1.

\appendix

\section*{Supplemental material}

\subsection{Exact tensor network for a Gibbs distribution}
It is possible to represent the Gibbs distribution exactly by a tensor network
\begin{equation}
P_E  (\zz) = \sum_{\iii} \prod_{\ell} A^{[\ell] z_\ell}_{\iii_{\ell}}, \nonumber
\end{equation}
where the bond dimensions are $\chi = d$.

The generality of such a representation follows from the arguments justifying the generality of the tensor network factorization of the partition function as, for example, appears in Ref.~\cite{Levin2007}. Here we give explicit expressions for the tensors $A^{[\ell]}$ corresponding to the two systems discussed in the main text: the one-dimensional (1D) open Ising chain and the two-dimensional (2D) periodic square Ising system, both with $d=2$.

For the 1D open Ising chain the elements of tensors $A^{[\ell]}$ are
\begin{align}
A^{[1] z_1}_{i_1} =& \exp \left[ \beta z_1 (J  i_1/2 + \lambda ) \right] ,  \nonumber \\
A^{[\ell] z_\ell}_{i_{\ell-1} i_\ell} =& \exp \left[\beta z_\ell (J  (i_{\ell-1} + i_\ell )/2 + \lambda ) \right], \; \; \; 1 < \ell < N ,  \nonumber\\
A^{[N] z_N}_{i_{N-1}} =& \exp \left[ \beta z_N (J  i_{N-1}/2 + \lambda ) \right] . \nonumber
\end{align}
The labeling of the indices $i_{\ell}$,  for $\ell = 1,\dots,N-1$, makes it clear which pairs are contracted. The indices take values $\pm 1$ and so the matrices $A^{[\ell] z_\ell}$ for $1<\ell< N$ are $2 \times 2$ (the boundary vectors $A^{[1] z_1}$ and $A^{[N] z_N}$ are $1 \times 2$ and $2 \times 1$, respectively).

For the 2D periodic square Ising system the tensors $A^{[\ell]}$ are identical and their elements are given by
\begin{align}
A^{[\ell] z_\ell}_{i_1 i_2 i_3 i_4} =& \exp \left[\beta z_\ell \left (J  \sum_{\ell'=1}^4 i_{\ell'}/2 + \lambda  \right ) \right] .  \nonumber 
\end{align}
The indices should be paired to form a square periodic lattice. The order of the indices at any one site is not important since the tensors are symmetric in the indices $i_1,\dots,i_4$. As before, the indices $i_{\ell'}$ for $\ell' = 1,\dots,4$ take values $\pm 1$.

\subsection{Time-dependence of the bias}
In our calculations the bias $\lambda (t)$ is varied between $\lambda_0$ and $\lambda_1$ according to a smoothed tanh ramp. The exact time-dependence used is
\begin{align}
\lambda(t) = \lambda_0 + \frac{\lambda_1 - \lambda_0}{2}\left [ 1+ \frac{ \tanh \left(  \sin \left[ \left(  \frac{t}{\tau} - \frac{1}{2} \right) \pi \right] \right)}{\tanh \left( 1  \right)} \right] , \nonumber
\end{align}
where $\tau$ is the duration of the quench.

\subsection{Single-time compression}
With the system out of equilibrium, we wish to determine the effect of compressing a distribution $P (\zz, t)$ (similarly $Q (\zz, t)$) at some time $t$. In other words we wish to construct a tensor network of small dimension $\chi$ that closely approximates the distribution
\begin{equation}
P (\zz, t) \approx \sum_{\iii} \prod_{\ell} A^{[\ell] z_\ell}_{\iii_{\ell}} . \nonumber
\end{equation}
For a 1D system with open boundaries the tensor network takes the matrix product form
\begin{equation}
 \sum_{\iii} \prod_{\ell} A^{[\ell] z_\ell}_{\iii_{\ell}} = A^{[1] z_1} A^{[2] z_2} \cdots A^{[N] z_N} , \nonumber
\end{equation}
with $\chi \times \chi$ matrices $A^{[\ell] z_\ell}$ for $1<\ell< N$, and $1 \times \chi$ and $\chi \times 1$ boundary vectors $A^{[1] z_1}$ and $A^{[N] z_N}$. It is well-known how to find an accurate matrix product approximation to a distribution $P (\zz, t)$ using singular value decomposition decimation, as explained in Refs.~\cite{Vidal2004,Johnson2010}. Note that the matrix product obtained may not be optimally accurate for the observable of interest, and thus even more effective compression may be possible than that we have reported. However, singular value decomposition decimation is closely related to how compression occurs in the operation of the time-evolution block decimation (TEBD) algorithm, and thus we expect the conclusions regarding compressibility at a single time here to be most relevant to repeated compression within the TEBD algorithm.

\subsection{Time-evolution block decimation}
TEBD is a standard method, detailed in Refs.~\cite{Vidal2004,Johnson2010}, for evolving a matrix product representation of some distribution $P (\zz, t)$ according to an equation of the form
\begin{align}
\pd{P (\zz, t)}{t} &= \sum_{\zz'} H  (\zz,\zz',t) P (\zz', t) . \nonumber
\end{align}
The algorithm proceeds by breaking the evolution into timesteps $\delta t$. Evolving the distribution $P (\zz, t)$ over a timestep takes it away from one expressed in terms of a matrix product of dimension $\chi$. Thus the distribution is repeatedly, within each timestep, re-compressed to this form using the singular value decomposition. The method we use to discretize time is a typical one and is explained in detail in Ref.~\cite{Johnson2010}. It is accurate up to errors that are second-order in $\delta t$. 

The most typical implementations of TEBD are compatible with two-site nearest-neighbor Hamiltonians only. Since the values of the transition rates appearing in Glauber dynamics depend on the energies at three nodes, the Hamiltonian is not naturally two-site.
To bring it into standard form, rather than $N$ single nodes with $d=2$, the system is broken down into $N/2$ supernodes with $d=4$ according to $\zz = ( z_{1,2}, \dots , z_{N-1,N} )$, where $z_{2\ell-1,2\ell}= \{ (-1,-1),(-1,1),(1,-1),(1,1) \}$ for $\ell = 1, \dots, N/2$. In the supernode picture the Hamiltonian is naturally formed of two-site nearest-neighbor terms.

Finally, the calculation results and scaling of resources quoted in the main text is for fixed IEEE 64-bit double precision computing. Though not necessary for the calculations presented here, higher than normal precision may be required for very large systems (see e.g.\ Ref.~\cite{Carlon1999}) due to the number of orders of magnitude spanned by the observable and probability distributions. 

\subsection{Tensor renormalization group}
We have used the tensor renormalization group method to calculate the sum $\sum_{\zz \in S} Q (\zz , t)$ for the case of an $N = 64 \times 64$ 2D periodic square Ising system, $O(t) = \ee^{-\beta W(t)}$ and $S$ the set of configurations with $z_1 = 1$. We are able to use tensor renormalization group methods because of the equilibrium structure $Q (\zz , t) = P_{t} (\zz) Z_t(\beta)/ Z_0(\beta)$ taken by the non-equilibrium distribution $Q (\zz , t)$ for this observable. 

To see this, we use the above to rewrite $\sum_{\zz \in S} Q (\zz , t) = (Z_t(\beta)/ Z_0(\beta)) \sum_{\zz \in S} P_{t} (\zz) $, where $P_t (\zz) $ is the equilibrium Gibbs distribution corresponding to the bias $\lambda(t)$ at time $t$, and $Z_t(\beta)$ the corresponding partition function. Thus the calculation reduces to calculating partition functions $Z_t(\beta)$ and partial sums $\sum_{\zz \in S} P_{t} (\zz) $ over Gibbs distributions conditioned upon the value of a single spin $z_1 = 1$. Given the representation of a Gibbs distribution $P_t  (\zz)$ by a tensor network, the partition function $Z_t = \sum_\zz P_t  (\zz) = \sum_{\iii} \prod_{\ell} T^{[\ell] }_{\iii_{\ell}}$ is merely a contraction of transfer tensors $T^{[\ell] } = \sum_{z_\ell} A^{[\ell] z_\ell}$ with a square geometry. The partial sum $\sum_{\zz \in S} P_{t} (\zz) = \sum_{\iii} \prod_{\ell} T^{[\ell] }_{\iii_{\ell}}$ is similar except with $T^{[1] } = A^{[\ell] z_1}$. 

These contractions can be performed using the well-known tensor renormalization group method~\cite{Levin2007}, the relevant details about which we provide here. The strategy is first to perform the contractions involving each set of four nodes making up a node of a coarse-grained square lattice. We are then left with tensor network in a square lattice with the number of nodes reduced by a factor of four, but the dimension of each index potentially increased by a power of $4$. Next, we compress this to obtain smaller index dimension $\chi$. The contraction/compression process is then repeated iteratively until a single node remains, and the contraction can be trivially evaluated. 

For $N = 64 \times 64$ this requires six iterations. We found it acceptable to use an intermediary dimension $\chi = 3$. 

\subsection{Dynamical Monte Carlo}
We have used dynamical Monte Carlo to calculate the sum $\sum_{\zz \in S} P (\zz , t)$ for the case of an $N = 64 \times 64$ periodic square Ising system and $S$ the set of configurations with $z_1 = 1$. We were able to use dynamical Monte Carlo because, by defining an observable $O(\zz)$ taking values $1$ if $\zz \in S$ and $0$ otherwise, we have $\sum_{\zz \in S} P (\zz , t) = \sum_{\zz} O(\zz) P (\zz , t)$ i.e.\ the quantity of interest is just the expected value $\langle O(t) \rangle$ of this observable.

We implement a variable-timestep dynamical Monte Carlo algorithm (as e.g.\ explained in Ref.~\cite{Jensen2003}) to sample $N_{\mathrm{s}}$ paths and estimate $\langle O(t) \rangle$ by averaging the values taken by the observable over such paths. Because the variance of the observable is necessarily less than unity, the expected error due to insufficient sampling must be less than $1/\sqrt{N_{\mathrm{s}}}$, which is small for the $N_{\mathrm{s}} = 14790$ we used.

The only source of error other than insufficient sampling is an approximation made when estimating the time until the next transition as part of the sampling of a path. We assumed for simplicity that the transition rates stayed constant between transitions, which ignores their varying due to the change in $\lambda(t)$. Since the average time between transitions is on the order of $(64^2 h)^{-1}$ and $\beta \lambda$ is varied by unity over a time $10 h^{-1}$ in our example, the approximation induced by this is negligible.

\end{document}